# Electron wave packet evolution under the influence of random electric field

D.I. Kulik[1], S.I. Pavlik[2], Yu.S. Oseledchik

*Department of Automated Systems of Process Control, Zaporizhzhya State Engineering Academy, 226 Soborny Ave., 69006 Zaporizhzhya, Ukraine.*

E-mail: sci.kulik@ya.ru[1], sipavlik@yahoo.com[2]

**Abstract.** We consider an electron wave packet evolution in the external deterministic electric field with Gaussian white noise addition. An explicit expression was given for the averaged probability flux with Gaussian initial wave function. We show that the noise addition increases wave packet spreading rate, which has a significant influence on flux amplitude and may play both destructive and constructive role. This depends on the form and amplitude of external deterministic field.

**Introduction**

The purpose of this paper is to consider the electron wave packet movement in the stochastic external field. The evolution of the initial state of particle in the quantum mechanics can be described by the time-dependent Schrodinger equation (TDSE). The obtained results are fully determined by the solution of correspondent partial differential equation. Different aspects of this problem have an important signification and considerable interest in terms of substantiation of quantum mechanics [1] and different applications [2]. Since problems like those are deeply analyzed in many textbooks in quantum mechanics, it is hard to except new results as, for example, for the evolution of electron wave packet in external fields. Although, if the evolution of a Gaussian wave packet in the TDSE is indeed an elementary textbook example, then the problem is far less understandable in case of random perturbations. The general class of problems for the quantum dynamics of a particle in a time dependent random (Markovian) potential has been treated in works [3,4]. Furthermore, the quantum evolution of the harmonic oscillator driven by time-dependent random electric field has been investigated in [5].

We would like to note that there has been growing interest in the study of quantum diffusion in a randomly fluctuating medium potential recently. For a dynamically disordered continuum, it was found that temporal fluctuations of random potential give rise to cubic growth of mean squared displacement with time [6]. Recently, it was considered a problem of evolution of the relativistic particle wave packet in a time-dependent random potential with $\delta$ correlation in time [7]. One more question that causes a considerable interest is the result of noise influence in the quantum system, including ultrashort pulse generation process. Generally, the influence of noise on the system can be classified as constructive (positive) and destructive (negative). In recent researches, positive possibilities of noise influence on the photoionization of atoms [8], high harmonics generation and attosecond pulse generation processes [9] were shown. Particularly, the enhancement of probability flux was observed.

Also, usage of molecular gases for generation of optical radiation has a significant interest. Molecules potentially have bigger nonlinear response and have many factors for its optimization, such as form, size, orientation, etc. However, it is some experimental difficulties connected with radiation obtaining in such mediums. As it was shown in [10], to transform molecules' radiation to the shortwave range, it must be in the excited state or in the process of started dissociation. At the same time, in the article [11] it was shown that the influence of stochastic component on the molecules can lead to their dissociation. It shows an undoubted interest to the researches related with the influence of noise on the ultrashort generation process on the atomic and molecular mediums.

Our research was initiated by the circumstance that for the interpretation of results in particle dynamics in external random potential with time dependence, it is necessary for the first to analyze the dynamics of electrons' wave packet. We consider the solution of TDSE with given wave function at the initial moment of time in the external electric field which is split on the determinant and random parts. We are interesting in time evolution of the probability flux which is averaged over external noise in given observation point. As the result, an explicit expression was given for the averaged flux with Gaussian initial wave function.

**1. The particle in the electric field**

It is well known that particle with the charge $-e$ in electric field $E(t)$ is described by the TDSE

$$i\hbar \frac{\partial \psi}{\partial t} = -\frac{\hbar^2}{2m}\frac{\partial^2 \psi}{\partial x^2} - eE(t)x\psi. \tag{1}$$

For simplicity, $\tau_0$ was chosen as the time unit and the $x_0$ as the spatial one to make transformations as follows

$$\frac{x}{x_0} \to x, \quad \frac{t}{\tau_0} \to t, \quad \frac{E}{E_0} \to E,$$

where $E_0 = \frac{\hbar}{ex_0\tau_0}$; $\frac{x_0^2}{\tau_0} = \frac{\hbar}{m}$. Thus, at the given replacement, there is only one parameter which was chosen arbitrarily. Particularly, if one uses Hartree atomic units system, $x_0 = \frac{4\pi\varepsilon_0\hbar^2}{m_e e^2} = 1$ $a.u.$ Thus Eq. (1) can be written as

$$i\frac{\partial \psi}{\partial t} = -\frac{1}{2}\frac{\partial^2 \psi}{\partial x^2} - E(t)x\psi. \tag{2}$$

The wave function corresponds to initial condition

$$\psi(x,0) = \varphi_0(x) \tag{3}$$

and was defined at the whole real axis. The Eq. (2) can be significantly simplified by applying the Fourier transform

$$\psi(x,t) = \int \frac{dk}{2\pi} \psi(k,t)e^{ikx}. \tag{4}$$

So we consider the Schrodinger equation in momentum space

$$i(\psi_t + E(t)\psi_k) = \frac{k^2}{2}\psi, \qquad (5)$$

with the initial condition

$$\psi(k,0) = \varphi_0(k). \qquad (6)$$

We applied the method of characteristics to solve the Eq. (5).

The left-hand side of the Eq. (5) is a total derivative of $\psi$ along the curve defined by equation

$$\frac{dk}{dt} = E(t). \qquad (7)$$

On integration we get

$$k = \int_0^t E(t')dt' + \xi = f(t) + \xi. \qquad (8)$$

For different values of $\xi$ we get a family of curves in the $(k,t)$ plane. Since $k = \xi$ at initial time $t = 0$, then $\psi(k,0) = \psi(\xi,0) = \varphi_0(\xi)$ is the initial condition from the equation

$$i\frac{d\psi}{dt} = \frac{1}{2}[f(t)+\xi]^2 \psi. \qquad (9)$$

On integration Eq. (9) and using Eq. (8) we have

$$\psi(k,t) = \varphi_0(k - f(t))e^{-\frac{i}{2}\int_0^t dt'(k+f(t')-f(t))^2}. \qquad (10)$$

Eq. (10) is an explicit solution, which can be used to calculate different physical values. In present work we interesting in the time evolution of probability flux at the predetermined spatial point

$$j(x,t) = \text{Im}(\psi^* \psi_x). \qquad (11)$$

The flux satisfies the continuity equation

$$\frac{\partial \rho}{\partial t} + \frac{\partial}{\partial x} j(x,t) = 0,$$

where $\rho = |\psi^2|$. Remind that the transformation to the dimensionless flux is given by

$$j(x,t) \to \frac{x_0^2}{\tau_0} j(x,t).$$

Performing the Fourier transform of Eq. (11) we easily find

$$j(q,t) = \int_{-\infty}^{\infty} \frac{dk}{2\pi} \psi^*\left(k - \frac{q}{2}, t\right) \psi\left(k + \frac{q}{2}, t\right) k. \tag{12}$$

Thus, using Eq. (10), the evolution of probability flux with specified initial condition is determined by Eq. (12).

Let us consider a simple example. Suppose that at the initial moment of time the wave function is a wave with the momentum $k_0$, and it extends in the positive direction of spatial axis $x$. Thus,

$$\varphi_0(x) \sim e^{ik_0 x}, \tag{13}$$

or

$$\varphi_0(k) \sim 2\pi\delta(k - k_0). \tag{13'}$$

Thus we can find, by substituting Eq. (13') to the Eq. (10)

$$\psi(k,t) = 2\pi\delta(k - k_0 - f(t)) e^{-\frac{i}{2}\int_0^t dt'[k + f(t') - f(t)]^2}. \tag{14}$$

Using Eq. (14) in the Eq. (12), we easily find

$$j(q,t) = 2\pi\delta(q)\left[\frac{q}{2} + k_0 + f(t)\right] e^{-i\int_0^t dt'\left[\frac{q^2}{2} + q(k_0 + f(t'))\right]}, \tag{15}$$

and then

$$j(t) = \int \frac{dq}{2\pi} j(q,t) e^{iqx} = k_0 + f(t). \tag{16}$$

The more interesting case can be obtained for nontrivial initial conditions. So let us consider that the wave function in initial moment of time is a Gaussian wave packet

$$\psi(x,0) = \varphi_0(x) = \frac{1}{\sqrt{\sigma\sqrt{\pi}}} e^{-\frac{x^2}{2\sigma^2}}. \tag{17}$$

The Fourier transform in momentum space

$$\varphi_0(k) = \sqrt{2\pi^{1/2}\sigma}\, e^{\frac{-k^2}{2\sigma^2}}. \tag{18}$$

Substituting Eq. (18) into Eq. (10) it is easily rewrite Eq. (12) for flux:

$$j(q,t) = \int \frac{dk}{2\pi}(k+f)\varphi_0^*\!\left(k-\frac{q}{2}\right)\varphi_0\!\left(k+\frac{q}{2}\right) e^{-iq\int_0^t dt'(k+f(t'))}, \tag{19}$$

where, using Eq.(18) we find

$$\varphi_0^*\!\left(k-\frac{q}{2}\right)\varphi_0\!\left(k+\frac{q}{2}\right) = 2\sqrt{\pi}\sigma\, e^{-\sigma^2\left[k^2+\frac{q^2}{4}\right]}. \tag{20}$$

Substituting Eq. (20) into Eq. (19), we obtain

$$j(x,t) = 2\sqrt{\pi}\sigma \int \frac{dk}{2\pi}[k+f(t)]e^{-\sigma^2 k^2} \int \frac{dq}{2\pi}\cdot e^{\frac{-\sigma^2 q^2}{4}+iqY(x,t)}, \tag{21}$$

where $Y(x,t) = x - kt - \Phi(t)$; for simplicity, it was set the designation $\Phi(t) = \int_0^t dt'\, f(t')$. The integrals in Eq. (21) become trivial for calculations using expression as follows:

$$\int \frac{dq}{2\pi} e^{iqy-aq^2}(A+Bq) = \frac{1}{\sqrt{4\pi a}} e^{\frac{-y^2}{4a}}\left[A + iB\frac{y}{2a}\right].$$

Thus, we find general expression for probability flux:

$$j(x,t) = \frac{1}{\sqrt{\pi\left(\sigma^2 + \frac{t^2}{\sigma^2}\right)}}\cdot e^{-\frac{(x-\Phi)^2}{\sigma^2 + \frac{t^2}{\sigma^2}}}\left[f(t) + \frac{x-\Phi}{\sigma^2 + \frac{t^2}{\sigma^2}}\left(\frac{t}{\sigma^2}\right)\right] \tag{22}$$

Let us consider the free wave packet, $f(t) = 0$. The distribution of probability flux in this case will be symmetrical about a point $x = 0$ while wave packet spreads at $x > 0$, $j > 0$ and at $x < 0$, $j < 0$. Thus at point $x = 0$ probability flux is always zero, $j = 0$. In the presence of constant electric field (as an example, in positive direction, $E > 0$), spreading of wave packet accompanies with the shift of every single modes. Wherein the divide point $j(x_0(t)) = 0$ shifts at the negative direction of the $x$ axis according to law

$$x_0(t) = \Phi - \frac{\sigma^4 + t^2}{t} f, \qquad t > 0.$$

For the constant field, $E = const$, $x_0(t) = -E\left(\sigma^4 + \frac{t^2}{2}\right)$.

Thus one expects that after some time the flux will be fully directed into the positive direction of the axis.

## 2. The influence of noise

Now let us consider how will be evolution changed after adding the additional noise term to the deterministic electric field.

For every realization of the external stochastic field, we will calculate the evolutional curve of flux which will be subsequently averaged with respect to specified distribution of random source.

We assume that stochastic electric field $\eta(t)$, which is included as additional term, $E(t) \to E(t) + \eta(t)$, is the Gaussian white noise with zero mean and has the $\delta$-function correlation,

$$\langle \eta(t) \rangle = 0; \qquad \langle \eta(t)\eta(t_1) \rangle = 2D\delta(t - t_1). \qquad (23)$$

Here the brackets $\langle (...) \rangle$ denote the ensemble for variance samples of the random electric field:

$$\langle (...) \rangle = \int D\eta (...) P[\eta(t)], \qquad (24)$$

where $P[\eta(t)]$ is the statistical weight functional of the random electric field:

$$P[\eta(t)] = Ce^{-\frac{1}{4D}\int dt' \eta^2(t')} \qquad (25)$$

with $C$ being the normalization constant.

Let us now consider the particle movement in an external electric field with noise addition treated in classical way. To describe this, we need to temporary return to the dimensional variables. As it is known, the equation of free electron motion in classical mechanics, with considering an additional noise term, can be written as follows

$$m\frac{d^2 x}{dt^2} = -eE - e\eta \qquad (26)$$

To find the statistical moments, which are described by the second-order differential equation, let us rewrite this equation as two first-order equations:

$$m\frac{dy}{dt} = -eE - e\eta, \quad \frac{dx}{dt} = y. \qquad (27)$$

Then we can easily find the equation for the first moment:

$$m\left\langle \frac{dy}{dt} \right\rangle = -e\langle E \rangle - e\langle \eta \rangle. \tag{28}$$

Thus, Eq. (28) is the same as the equation without noise addition, since $\langle \eta \rangle = 0$. Considering $<x(t)>$ as the center of electron wave packet, its trajectory is fully determined by the form of the external field.

Multiplying the first Eq. in (27) by $y$ and averaging the result, one can find the differential equation for an average kinetic energy:

$$\frac{d}{dt} < \frac{1}{2} my^2 > = -e<y>E - e<y\eta>. \tag{29}$$

To perform averaging over the realization of stochastic electric field $<y\eta>$, it can be used Furutsu-Novikov equation [12,13]. Thus we have

$$<y(t)\eta(t)> = \int_0^t <\eta(t)\eta(t')> \langle \frac{\delta y(t)}{\delta \eta(t')} \rangle dt' = -\frac{e}{m} D \tag{30}$$

From the Eq. (30), it follows that in classical case the noise addition has no influence on the average trajectory, but it is cause to increase of kinetic energy of the system. Particularly, from the Eq. (28) it follows that external noise does not change the flux.

Now let as turn back to the system, which was described in Chapter 1 and try to estimate the noise influence on it. To add a noise term to Eq. (22), we need to make next replacement:

$$f \to f_0 + \int_0^t \eta(t')dt' = f_0 + f(t);$$

$$\Phi \to \Phi_0 + \int_0^t f(t')dt' = \Phi_0 + \Phi(t), \tag{31}$$

where $f_0, \Phi_0$ - deterministic field terms and $\Phi(t)$ - field term that is bounded with stochastic influence. Besides, $f(t)$ is the derivative of the $\Phi(t)$: $f(t) = \Phi_t$. After the substituting Eq. (31) into Eq. (22), it needs to be averaged by the external noise. It can be done using Eq. (19), which can be rewritten as

$$j(x,t) = 2\sqrt{\pi}\sigma \int_{-\infty}^{\infty} \frac{dk}{2\pi}(k + f_0 + f)e^{-\sigma^2 k^2} \int_{-\infty}^{\infty} \frac{dq}{2\pi} e^{\frac{-\sigma^2 q^2}{4} + iqY_0 - iq\Phi(t)}, \tag{32}$$

where $Y_0(x,t) = x - kt - \Phi_0(t)$.

As is seen, averaging Eq.(32) according to Eq. (24), there are appear expressions of two types:

$$\langle e^{-iq\Phi} \rangle = e^{-\frac{q^2}{2}\langle \Phi^2 \rangle}, \tag{33}$$

$$\langle fe^{-iq\Phi} \rangle = -iq \langle \Phi f \rangle e^{-\frac{q^2}{2}\langle \Phi^2 \rangle} \tag{34}$$

and as it follows from Eq. (23)

$$\langle f(t_1)f(t_2) \rangle = 2D \min(t_1, t_2). \tag{35}$$

Thus after the averaging, using Eq. (33), Eq. (34) and Eq. (35), we easily find

$$\langle j(x,t) \rangle = 2\sqrt{\pi}\sigma \int_{k,q} e^{-\sigma^2 k^2 - \frac{\sigma^2}{4}q^2 + iqY_0} \left[ (k+f_0)\langle e^{-iq\Phi} \rangle + \langle fe^{-iq\Phi} \rangle \right]$$

$$= 2\sqrt{\pi}\sigma \int_{k,q} e^{-\sigma^2 k^2 - \left(\frac{\sigma^2}{4} + \frac{\langle \Phi \rangle^2}{2}\right)q^2 + iqY_0} \left[ (k+f_0) - iq\frac{\langle \Phi \rangle_t^2}{2} \right]. \tag{36}$$

After the integration by $k$ and $q$, we get next equation:

$$\langle j(x,t) \rangle = \frac{1}{\sqrt{\pi\left(s^2 + \frac{t^2}{\sigma^2}\right)}} e^{-\frac{(x-\Phi_0)^2}{s^2 + \frac{t^2}{\sigma^2}}} \left[ f_0 + \frac{x-\Phi_0}{s^2 + \frac{t^2}{\sigma^2}}\left(\frac{t}{\sigma^2} + \langle \Phi \rangle_t^2\right) \right], \tag{37}$$

where $s^2 = \sigma^2 + 2\langle \Phi^2 \rangle$; $\langle \Phi^2 \rangle = 2D\frac{t^3}{3}$; $\langle \Phi^2 \rangle_t = 2Dt^2$.

Comparing Eq. (37) with corresponding expression for flux without noise influence accounting (Eq. 22), it is seen, that the influence of noise doesn't lead to significant changes in analytical expression. At the same time, the rate of wave packet width change is increased. On Figure 1 there are shown flux dependences from time at the observation point $x = 20$ for different noise intensities in the constant electric field. As is it seen, the increasing of the noise intensity leads to decreasing the amplitude of flux curve, but also it causes the widening of curve in time. In other words, the noise leads to additional spreading of wave packet.

Consider the alternating electric field (atomic units will be used). It was chosen to set the field as the ultrashort femtosecond pulse with the form

$$E(t) = E_0 \cdot e^{-5(\omega t/2\pi - 1)^4} \sin \omega t, \tag{38}$$

which is used in work [10] to evaluate high harmonics generation and attosecond radiation bursts from the molecular medium due to returning of wave packet to the origin including the influence of molecular potential. Here the amplitude of pulse was chosen as $E_0 = 0.1$ and the frequency as $\omega = 0.114$ ($\lambda \sim 400\ nm$). The flux dependencies from time are shown on Figure 2. Using this parameters we can show the possibility of stochastic enhancement of the flux at observed spatial point $x = 20$. There are two influences: the weak part of laser field, which corresponds to the whole wave packet shift and the stochastic force; this parameter increases the spreading. Thus, at the period of time from 0 to 30, when the field is weak, one can see the spreading of wave packet goes faster than it moves, so it reaches point $x$ earlier and the flux has bigger absolute value. But after time point 30, the enhancement effect vanishes and amplitude decreases due to increasing of field amplitude, so the wave packet shift rate prevails over its spreading.

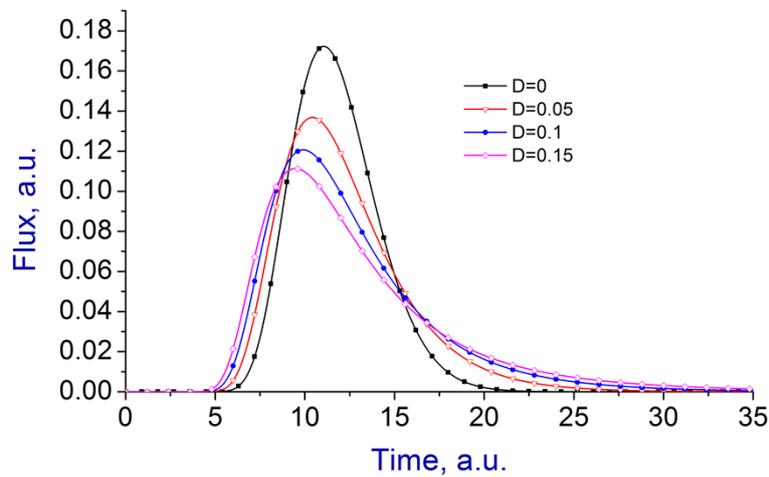

**Figure 1.** The probability flux (left axis, solid line) at $x = 20$ as a function of time calculated for the different noise intensity $D$ in constant electric field $E = 0.3$

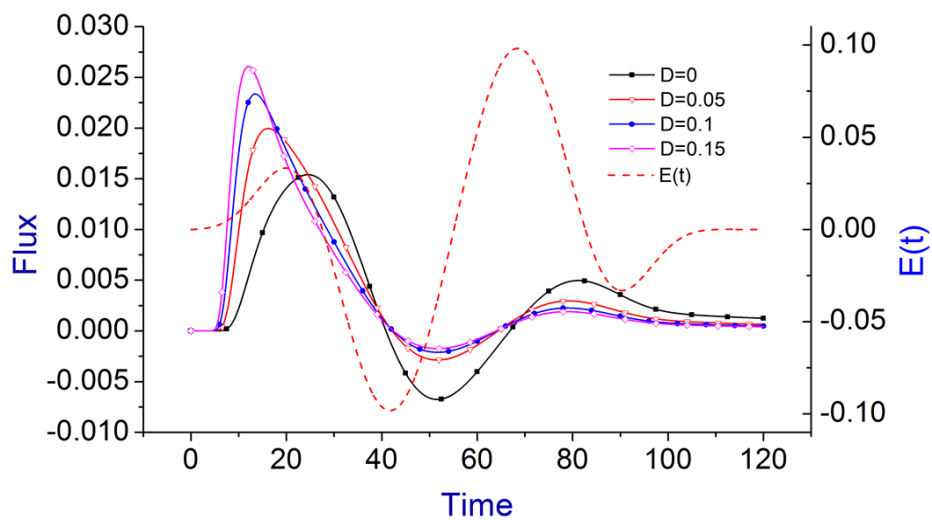

**Figure 2.** The probability flux (left axis, solid line) at $x = 20$ as a function of time calculated for the different noise intensity $D$ in the electric field of femtosecond laser pulse $E(t)$ (right axis, dotted line)

**Conclusions**

We have researched the time-dependent evolution of the particle wave function in external field which was considered as sum of deterministic and random terms. It was given an explicit solution for average probability flux. Two cases there were investigated. In the first one, the deterministic part of field was constant and at the second it was chosen as a femtosecond laser pulse with wavelength ~ 400 *nm* (the second harmonic from the Ti:Sa laser radiation). It was shown that the noise addition to the system leads to the additional spreading of wave packet in constant field and in pulse field both. However, the dependence of flux from time for both cases is different: the flux amplitude decreases at the constant field case but in pulse field case it can be seen the region with increasing of flux. Thus, in observe system, the noise can play both constructive and destructive role which depends on the form of external field.